\documentclass[mathleft
]{an}
\usepackage{graphicx}
\usepackage{times}
\begin{document}

% The following seven commands are intended for editorial usage and should be ignored by
% the author(s).
\Pagespan{789}{}% Document's page range. 
% If second parameter is left empty, the last page is computed automatically.
\Yearpublication{2006}%
\Yearsubmission{2005}%
\Month{11}%   
\Volume{999}%  
\Issue{88}% 
% \DOI{This.is/not.aDOI}% 

\title{The X-ray--Infrared/Submillimetre Connection and the \\
       Legacy Era of Cosmology}

\author{D.~M.~Alexander\inst{1}\fnmsep\thanks{Corresponding author:
  \email{d.m.alexander@durham.ac.uk}\newline}
%Example 
%for footnote, note the usage of the \texttt{fnmsep}
%command as separator between institute number and footnote mark} 
%\and  G.H. Ostwriter\inst{2,3}
}
\titlerunning{The X-ray--infrared/submillimetre Connection}
\authorrunning{D.~M.~Alexander}
\institute{
$^1$ Department of Physics,
Durham University,
South Road,
Durham,
DH1 3LE, UK}

\received{30 May 2005}
\accepted{11 Nov 2005}
\publonline{later}


\abstract{% 
We review some recent results on the identification and
characterisation of Active Galactic Nuclei (AGN) obtained by cross
correlating X-ray surveys with infrared and submillimetre surveys. We
also look toward the scientific gains that could be achieved from an
{\it XMM-Newton} survey of the medium-deep legacy fields
that are being observed at $\approx$~1--850~$\mu$m.  }

\maketitle

\section{Introduction}

Deep infrared (IR) and submillimetre (submm) surveys are providing a
sensitive view of the Universe at $\approx$~3--850~$\mu$m (e.g.,\
Coppin et~al. 2006; Dole et~al. 2006; Frayer et~al. 2006), revealing
large numbers of dust-obscured starburst galaxies and Active Galactic
Nuclei (AGN). These surveys have found that luminous IR galaxies
strongly evolve out to $z\approx$~1--2, implying that a large
(possibly dominant) fraction of the growth of galaxies occurred at
high redshift in an IR-bright phase. Detailed studies have, indeed,
suggested that at least half of all newly born stars are formed in
LIRGs ($L_{\rm IR}\approx10^{11}$--$10^{12}$~$L_{\odot}$) hosted in
moderate-mass galaxies at $z<1.5$ (e.g.,\ Le Floc'h et~al. 2005;
Perez-Gonzalez et~al. 2005), and that distant ULIRGs ($L_{\rm
IR}>10^{12}$~$L_{\odot}$) represent a key phase in the formation of
the spheroids of today's massive galaxies (e.g.,\ Chapman et~al. 2005;
Swinbank et~al. 2006).

Although providing detailed insight into the formation and evolution
of the majority of the stellar population, a large uncertainty in the
interpretation of these studies is the contribution (or
``contamination'') from IR-bright AGNs, which could be
significant. For example, the best estimates on the total amount of
star formation overproduce the stellar-mass density by a factor of
$\approx$~2 (e.g.,\ Chary \& Elbaz 2001; Hopkins \& Beacom 2006), a
discrepancy which would be neatly explained by a large population of
hitherto unidentified obscured AGNs (potentially Compton-thick
objects; see Comastri 2004 for a review). Furthermore, given the tight
relationship between the mass of galaxy spheroids and their central
black holes (e.g.,\ Magorrian et~al. 1998; Tremaine et~al. 2002), it
might also be expected that any major star-formation phase co-incides
with periods of AGN activity, during which the black hole is grown in
tandem. Although the global evolution of star formation and AGN
activity is comparatively well constrained (e.g.,\ Madau et~al. 1996;
Barger et~al. 2005; Hasinger et~al. 2005; Hopkins \& Beacom 2006), the
details of how galaxies and their black holes grew (e.g.,\ as a
function of environment, cosmic epoch, and mass) are still largely
unknown.

Arguably, the most direct indication of AGN activity is the detection
of luminous hard X-ray emission (i.e.,\ $>$~2~keV). Hard X-ray
emission appears to be a universal property of AGNs, giving a direct
``window'' on the emission regions closest to the black hole (e.g.,\
Mushotzky, Done, \& Pounds 1993), and it can provide a secure AGN
identification in sources where the optical signatures and
counterparts are weak or even non existent (e.g.,\ Alexander et~al.
2001a; Comastri et~al. 2002). Hard X-ray emission is also relatively
insensitive to obscuration (at least for sources that are Compton
thin; i.e.,\ $N_{\rm H}<1.5\times10^{24}$~cm$^{-2}$) and any hard
X-ray emission from star formation in the host galaxy is often
insignificant when compared to that produced by the AGN. Importantly,
the X-ray emission provides a direct measurement of the
primary power of the AGN, crucial information when estimating mass
accretion rates and the relative bolometric contributions from AGN and
star-formation activity (e.g.,\ Alexander et~al. 2005a,b). However, 
in the case of Compton-thick AGNs, where the observed X-ray emission
is often very faint and dominated by reflected/scattered components,
the identification of luminous mid-IR emission can provide a route to
estimating the intrinsic power of the AGN. These estimates can be
particularly accurate when optical and mid-IR spectroscopy are
available (e.g.,\ Lutz et~al. 2004).

Here we review some recent results obtained from the cross correlation
of X-ray and IR/submm surveys. We also look toward the scientific
gains that could be achieved from an {\it XMM-Newton} survey of the
legacy fields being observed at 1--850~$\mu$m
(VISTA, UKIDSS, {\it Spitzer}, {\it Herschel}, SCUBA2), the data of
which are typically made publicly available on short timescales.

\section{The Advances Made by {\it XMM-Newton} and {\it Chandra}}

Early studies on the connection between X-ray and IR/submm sources
focused on nearby objects, or rare, luminous AGNs (e.g.,\ Barcons
et~al. 1995; Alexander et~al. 2001b). The orders of magnitude increase
in sensitivity that {\it XMM-Newton} and {\it Chandra} have provided
over previous X-ray observatories, allied to the sensitivity gains
made by {\it Spitzer} and premier submm instruments (e.g.,\ SCUBA and
MAMBO), can now yield insight into the role of typical AGNs in
IR/submm galaxies out to high redshift.

\begin{figure}
\includegraphics[width=80mm]{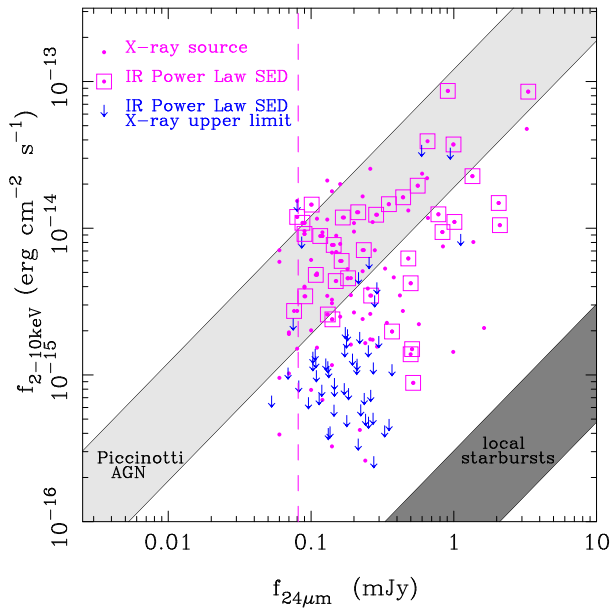}
\caption{Hard X-ray flux versus 24~$\mu$m flux density for X-ray
identified (filled dots) and IR-identified AGNs (open squares and
arrows) in the {\it Chandra} Deep Field-South. The light shaded region
indicates where (comparatively unobscured) AGNs and starbursts with
flux ratios consistent with local objects would lie. Taken from
Alonso-Herrero et~al. (2006).}
\label{label1}
\end{figure}

The current X-ray--IR cross-correlation studies have typically focused
on characterising the IR Spectral Energy Distributions (SEDs) of X-ray
and IR selected AGNs. As known since IRAS, about half of the nearby
AGN population are identifiable as AGNs from their IR SEDs while the
other half have IR properties more consistent with those expected from
star formation (see Fig.~3 of Alexander 2001). A good example of the
latter is NGC~6240, a nearby powerful AGN hosted in a starburst
galaxy, that only clearly reveals the signatures of luminous AGN
activity at hard X-ray energies (e.g.,\ Vignati et~al. 1999). A
similar dichotomy is being found for more distant objects identified
in {\it Spitzer} surveys. For example, using only moderately deep
X-ray observations, up-to $\approx$~30\% of the X-ray identified AGN
have starburst-like IR SEDs (e.g.,\ Franceschini et~al. 2005; Polletta
et~al. 2007), and the fraction approaches $\approx$~50\% with deeper X-ray
data (e.g.,\ Alonso-Herrero et~al. 2004; Alexander 2006; see
Fig.~1). Conversely, selecting IR galaxies with AGN-like SEDs, it has
also been possible to identify heavily obscured (potentially Compton
thick) AGNs that are not detected at X-ray energies (e.g.,\
Alonso-Herrero et~al. 2006; Polletta et~al. 2006; Donley
et~al. 2007). The latest results suggest that Compton-thick AGN in the
distant Universe could be wide spread, accounting for a large fraction
of the growth of black holes (e.g.,\ Daddi et~al. 2007; Fiore
et~al. 2008; Alexander et~al. 2008b). 
It is only from a {\it combination} of X-ray and IR observations
that Compton-thick AGNs can be effectively identified; radio and X-ray
observations can also provide a sensitive probe of Compton-thick AGNs
(e.g.,\ Donley et~al. 2005; Mart{\'{\i}}nez-Sansigre
et~al. 2007) and will become key resources with LOFAR and the SKA.

Cross correlation studies of X-ray and submm surveys have shown that
submm-emitting galaxies (SMGs) are, typically, only detected in deep
X-ray surveys (see Fig.~1 of Alexander et~al. 2003). There are two
major reasons for this (1) the flat selection function of submm
blank-field surveys means that most SMGs are detected at 
high redshift ($z\approx$~2; e.g.,\ Smail
et~al. 2002; Chapman et~al. 2005), making them faint at most other
wavelengths, and (2) detailed X-ray spectral analyses have indicated
that the majority of the AGNs are heavily obscured and only moderately
luminous at X-ray energies (e.g.,\ Alexander et~al. 2005a). Since SMGs
are amongst the most bolometrically luminous galaxies in the Universe,
the relatively weak X-ray emission implies that the AGN activity is
unlikely to dominate the global energetics (e.g.,\ Alexander
et~al. 2005a). Indeed, mid-IR spectroscopy of SMGs has confirmed that the
dominant power source of SMGs is star formation (e.g.,\
Valiante et~al. 2007; Men{\'e}ndez-Delmestre et~al. 2007; Pope
et~al. 2008). However, the large fraction of SMGs that host AGN
activity ($\approx$~28--50\%) indicates that their black holes are
growing almost continuously throughout periods of intense star
formation (e.g.,\ Alexander et~al. 2005b). Careful assessment of the
black-hole and host-galaxy masses of SMGs indicates that their black
holes are smaller than those expected for comparably massive galaxies
in the local Universe (e.g.,\ Borys et~al. 2005; Alexander
et~al. 2008a), indicating that they must undergo an intense black-hole
growth phase before the present day. Observational evidence suggests
that this rapid black-hole growth phase is associated with optically
luminous quasar activity (e.g.,\ Page et~al. 2004; Stevens
et~al. 2005; Alexander et~al. 2008a), in general agreement with
predictions from simulations of SMG-like systems (e.g.,\ Granato
et~al. 2006; Chakrabarti et~al. 2007).

The combination of X-ray and IR/submm surveys have furthered our
understanding of AGNs in the distant Universe and increased the global
``census'' of AGN activity. However, there has
been little research on what causes distant black holes to
grow: for example, is it a function of local environment (i.e.,\
nearby/interacting galaxies) and/or large-scale structure environment (i.e.,\
clusters versus field)? To fully explore this 
issue over a representative range of environments requires sensitive
multi-wavelength coverage over large {\it contiguous} areas of the sky.

\section{The Legacy Era of Cosmology}

\begin{figure}
\includegraphics[width=80mm]{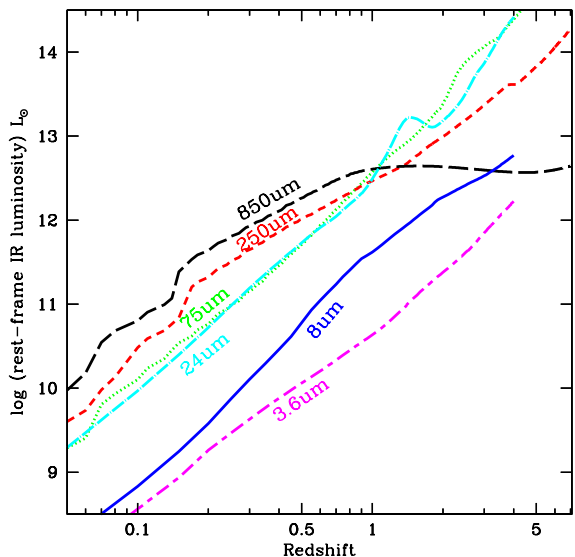}
\caption{Predicted luminosity at 8--1000~$\mu$m versus redshift for
some of the surveys being performed in the SWIRE legacy fields; the
luminosities are calculated using the Chary \& Elbaz (2001) SEDs. The
different curves represent different selection wavebands (as
annotated). Tracks used in figure produced by D.~Elbaz.}
\label{label1}
\end{figure}

Over the last few years, a large amount of astronomical resources have
been committed to compiling deep multi-wavelength data over large
areas of the sky. These extensive datasets provide the potential to
trace the growth of galaxies, their black holes, and the large-scale
structure that they reside in, across a large fraction of cosmic
time. To maximise their scientific potential, these observations are typically
made publicly available on short timescales.

In terms of observational cosmology, the {\it Spitzer}-SWIRE survey
probes a key region of sensitivity--solid angle parameter space at
$\approx$~3--160~$\mu$m (Lonsdale et~al. 2003, 2004). Covering
$\approx$~50~deg$^2$ over 6 fields, this survey traces the growth of
galaxies out to $z\approx$~1 across a broad range of environments,
from galaxy voids to galaxy superclusters (linear scales of
$\approx$~50--100~Mpc at $z\approx$~1). Upcoming surveys with {\it
Herschel} and SCUBA2 extend this coverage out to $\approx$~850~$\mu$m,
while surveys with VISTA and UKIDSS cover the shorter wavelength range
at $\approx$~1--2.5~$\mu$m.\footnote{See
http://astronomy.sussex.ac.uk/$\sim$sjo/Hermes/ for more details on
these surveys} In Fig.~2
we show the predicted IR sensitivity for some of these surveys; LIRGs
are detectable out to $z\approx$~0.5--2 and ULIRGs will be identified
out to higher redshifts. However, a major wavelength component missing
from this large legacy survey is complete and sensitive X-ray
coverage.

\subsection{The {\it XMM-Newton} Wide-Deep Survey:}

A moderately deep {\it XMM-Newton} survey of the $\approx$~10~deg$^2$
of the SWIRE fields, where there is the maximum overlap between the
different multi-wavelength surveys, would provide the most direct
constraints on the energetics of AGN activity and black-hole growth in
these fields. Crucially, such a large-area sensitive X-ray survey would
have sufficient source statistics, cover a wide-enough
range of environments and map out a large-enough volume, to constrain
the conditions in the distant
Universe that led to black-hole growth. This survey would require
sufficient sensitivity to be able to detect Seyfert galaxies at
$z\approx$~0.7--1 and quasars at $z\approx$~2--3, in the comparatively
obscuration-independent 2--10~keV band; this would
require {\it XMM-Newton} exposures of $\approx$~50~ks (i.e., $L_{\rm
  X}\approx10^{43}$~erg~s$^{-1}$ at $z\approx$~1 and $L_{\rm
  X}\approx10^{44}$~erg~s$^{-1}$ at $z\approx$~3). The {\it
  XMM-Newton} COSMOS survey (Hasinger et~al. 2007) provides the 
largest field available with
sufficient sensitivity to perform this experiment. However, it does not
cover all large-scale structure environments and the source statistics are too
poor to explore AGN activity as a function of different parameters. For
example, assuming 10 galaxy density bins (to measure the local
environment), 5 redshift bins, 5 X-ray luminosity bins, 2
large-scale structure bins (clusters versus non-clusters), and
10--20 objects per bin (for basic statistical constraints), requires a sample of
up-to 10,000 AGNs; on the basis of current number counts, this
could be achieved with a 50~ks survey of $\approx$~10~deg$^2$. We
briefly explore some of the scientific goals that this {\it XMM-Newton}
Wide-Deep survey could explore.

\noindent {\bf What makes black holes grow?}
Many theoretical models predict that black holes hosted in dense regions will
grow more rapidly than those hosted in low-density regions. The {\it
  XMM-Newton} Wide-Deep survey would be able to explore 
this as a function of many
different parameters, including investigating whether the environments
of the dominant class of AGN at a given redshift are different to those
found for the overall AGN population (i.e.,\ exploring what
is driving the so-called AGN ``cosmic downsizing''; e.g.,\ Barger
et~al. 2005; Hasinger et~al. 2005). This might reveal
that there are characteristic densities at which AGN activity is
triggered as a function of redshift, which would be key ingredients for
galaxy--AGN formation and evolution models. The exploration of
IR-selected galaxies at $z\approx$~1 has already indicated that galaxies in
dense regions are growing more rapidly than those in underdense
regions, in stark contrast to that found in the local Universe (e.g.,\
Elbaz et~al. 2007), and the {\it XMM-Newton} Wide-Deep survey would reveal whether
black-hole growth is initiated in the same type of environments.

\noindent {\bf Lifting the veil of the dust-obscured Universe:} 
The {\it XMM-Newton} Wide-Deep survey will provide efficient
identification of the presence of even heavily obscured AGNs in the
$\approx$~300,000
{\it Spitzer}-IRAC and $\approx$~100,000 {\it Spitzer}-MIPS sources
over the 10~deg$^2$ region, yielding a direct measurement of\
 the intrinsic luminosity of any AGN activity; $\approx$~3,000-6,000 X-ray
AGNs should be detected by {\it Spitzer}, including $\approx$~500 X-ray
obscured quasars. Due to the negative
K-correction at submm wavelengths, the SCUBA2 observations will
provide a redshift-independent view of the Universe and a route
to identifying the overdense environments in which massive
galaxies are forming. These often appear to be coincident with X-ray
absorbed quasar
activity (e.g.,\ Stevens et~al. 2004), and the {\it XMM-Newton}
Wide-Deep survey should detect $\approx$~100--200 such systems that
could be explored in detail (both
the quasar and any companion star-forming galaxies).

The large-area coverage of the {\it XMM-Newton} Wide-Deep survey would
also allow, for example,
constraints on the clustering of AGNs as a function of different
parameter space, the detection of high-redshift AGNs ($\approx$~500
objects are expected at $z>3$, which will be $\approx$~2--7 mags
fainter than those found by the SDSS),
the detection of rare luminous objects not found in
smaller fields (e.g.,\ luminous obscured submm-emitting quasars; $z>1$
galaxy clusters), the identification of $\approx$~100--150 luminous Compton-thick AGNs,
and could place detailed constraints on the evolution of AGN activity
in galaxy clusters out to $z\approx$~1.

\noindent {\bf Experimental Design:} A single 10~deg$^2$
field would be more biased than multiple large
fields (see \S3.3 of Oliver et~al. 2000) and a
10~deg$^2$ survey composed of a large number of small fields would not
allow for the efficient identification of large-scale structure
environments (which require a degree-sized field to be identified); the
latter point is also a reason why the {\it XMM-Newton} serendipity survey
cannot achieve the main science goal (the {\it XMM-Newton}
serendipity survey also does not have sufficient multi-wavelength coverage). A
more complete distribution of environments would be achieved from five
$\approx$~2~deg$^2$ fields. Taking into account of the multiwavelength
coverage and the visibility of fields to {\it XMM-Newton}, the
following five fields are optimal: COSMOS, CDF-S, XMM-LSS, XMM-Lockman
Hole, and ELAIS S1. At present these fields have 2.9~deg$^2$ of 
sufficiently sensitive X-ray coverage (including COSMOS) and 2~deg$^2$
of the XMM-LSS has $\approx$~20~ks coverage and would therefore only need to be
supplemented with $\approx$~30~ks {\it XMM-Newton}
observations. Assuming the tiling strategy of XMM-LSS, in order to
achieve uniform {\it XMM-Newton} coverage, the {\it XMM-Newton}
Wide-Deep survey could be realised with an additional 3~Ms of {\it
  XMM-Newton} exposure. 

The {\it XMM-Newton} Wide-Deep survey outlined above requires a large
allocation of the available {\it XMM-Newton} time. However, when
compared to the $\approx$~10+~Ms investment made at
$\approx$~1--850~$\mu$m 
in these fields
(not including the extensive optical and radio observations),
it is cost effective and would
provide a major astronomical resource
for decades to come, achieving sensitivity limits at least an order of
magnitude deeper than planned large-area hard X-ray surveys (e.g.,\
{\it e-ROSITA}). Furthermore, the
wide distribution of these fields across the
sky would allow for convenient scheduling of the {\it XMM-Newton}
observations, minimising the impact on other {\it XMM-Newton}
programs.

\acknowledgements I thank the Royal Society for support, and
A.~Alonso-Herrero and D.~Elbaz for providing data. The
motivation for the {\it XMM-Newton} legacy survey arose from
discussions with F.~Bauer, N.~Brandt, A.~Edge, M.~Jarvis, B.~Lehmer,
K.~Nandra, M.~Page, K.~Romer, C.~Simpson, I.~Smail, and M.~Watson.

%\newpage%%%%%%%%%%%%%%%%%%%%%%%%%%%%%%%%%%%%%%%%%%%%%%%%%%%%%%


\begin{thebibliography}{}
\bibitem[Alexander(2001)]{2001MNRAS.320L..15A} Alexander, D.~M.\ 2001, 
\mnras, 320, L15 
\bibitem[Alexander et al.(2001a)]{2001AJ....122.2156A} Alexander,
  D.~M., et~al.\ 2001a, \aj, 122, 2156 
\bibitem[Alexander et al.(2001b)]{2001ApJ...554...18A} Alexander, D.~M., et 
al.\ 2001b, \apj, 554, 18 
\bibitem[Alexander et al.(2003)]{2003AJ....125..383A} Alexander, D.~M., et 
al.\ 2003, \aj, 125, 383 
\bibitem[Alexander et al.(2005a)]{2005ApJ...632..736A} Alexander, D.~M., 
et~al.\ 2005a, \apj, 632, 736 
\bibitem[Alexander et al.(2005b)]{2005Natur.434..738A} Alexander, D.~M., 
et~al.\ 2005b, Nature, 434, 738 
\bibitem[Alexander(2006)]{2006astro.ph.12497A} Alexander, D.~M.\ 2006, 
ArXiv Astrophysics e-prints, arXiv:astro-ph/0612497 
\bibitem[Alexander(2008)]{2008submitted} Alexander, D.~M., et~al.\ 2008a, 
\aj, submitted
\bibitem[Alexander(2008a)]{2008submitted} Alexander, D.~M., et~al.\ 2008b, 
\apj, submitted
\bibitem[Alonso-Herrero et al.(2004)]{2004ApJS..154..155A} Alonso-Herrero, 
A., et al.\ 2004, \apjs, 154, 155 
\bibitem[Alonso-Herrero et al.(2006)]{2006ApJ...640..167A} Alonso-Herrero, 
A., et al.\ 2006, \apj, 640, 167 
\bibitem[Barcons et al.(1995)]{1995ApJ...455..480B} Barcons, X., et~al.\ 1995, \apj, 455, 
480 
\bibitem[Barger et al.(2005)]{2005AJ....129..578B} Barger, A.~J., et~al.\ 2005, \aj, 129, 578 
\bibitem[Borys et al.(2005)]{2005ApJ...635..853B} Borys, C., et~al.\ 2005, 
\apj, 635, 853 
\bibitem[Chakrabarti et al.(2006)]{2006astro.ph.10860C} Chakrabarti, S., 
et~al.\ 2006, ApJ, submitted (astro-ph/0610860) 
\bibitem[Chapman et al.(2005)]{2005ApJ...622..772C} Chapman, S.~C., et~al.\ 2005, \apj, 622, 772
\bibitem[Chary \& Elbaz(2001)]{2001ApJ...556..562C} Chary, R., \& Elbaz, 
D.\ 2001, \apj, 556, 562 
\bibitem[Comastri(2004)]{2004ASSL..308..245C} Comastri, A.\ 2004, 
Supermassive Black Holes in the Distant Universe, 308, 245 
\bibitem[Comastri et al.(2002)]{2002ApJ...571..771C} Comastri, A., et al.\ 
2002, \apj, 571, 771 
\bibitem[Coppin et al.(2006)]{2006MNRAS.372.1621C} Coppin, K., et al.\
2006, \mnras, 372, 1621
\bibitem[Daddi et al.(2007)]{2007arXiv0705.2832D} Daddi, E., et al.\ 2007, 
ApJ, 670, 173
\bibitem[Dole et al.(2006)]{2006A&A...451..417D} Dole, H., et al.\ 2006, 
A\&A, 451, 417 
\bibitem[Donley et al.(2005)]{2005ApJ...634..169D} Donley, J.~L., et~al.\ 2005, \apj, 634, 169 
\bibitem[Donley et al.(2007)]{2007ApJ...660..167D} Donley, J.~L., et~al.\ 2007, \apj, 660, 167 
\bibitem[Elbaz et al.(2007)]{2007A&A...468...33E} Elbaz, D., et al.\
2007, A\&A, 468, 33
\bibitem[Fiore et al.(2007)]{2007arXiv0705.2864F} Fiore, F., et al.\ 2008, 
ApJ, 672, 94
\bibitem[Franceschini et al.(2005)]{2005AJ....129.2074F} Franceschini, A., 
et al.\ 2005, \aj, 129, 2074 
\bibitem[Frayer et al.(2006)]{2006ApJ...647L...9F} Frayer, D.~T., et al.\ 
2006, \apjl, 647, L9 
%\bibitem[Geach et al.(2006)]{2006ApJ...649..661G} Geach, J.~E., et al.\ 
%2006, \apj, 649, 661 
\bibitem[Granato et al.(2006)]{2006MNRAS.368L..72G} Granato, G.~L., et~al.\ 2006, \mnras, 368, 
L72 
\bibitem[Hasinger et al.(2005)]{2005A&A...441..417H} Hasinger, G.,
Miyaji, T., \& Schmidt, M.\ 2005, A\&A, 441, 417
\bibitem[Hasinger et al.(2007)]{2007ApJS..172...29H} Hasinger, G., et
  al.\ 2007, \apjs, 172, 29 
%\bibitem[Heisler et al.(1997)]{1997Natur.385..700H} Heisler, C.~A., 
%Lumsden, S.~L., \& Bailey, J.~A.\ 1997, Nature, 385, 700 
\bibitem[Hopkins \& Beacom(2006)]{2006ApJ...651..142H} Hopkins, A.~M., \& 
Beacom, J.~F.\ 2006, \apj, 651, 142 
\bibitem[Le Floc'h et al.(2005)]{2005ApJ...632..169L} Le Floc'h, E., et 
al.\ 2005, \apj, 632, 169 
\bibitem[Lonsdale et al.(2003)]{2003PASP..115..897L} Lonsdale, C.~J., et 
al.\ 2003, \pasp, 115, 897 
\bibitem[Lonsdale et al.(2004)]{2004ApJS..154...54L} Lonsdale, C., et al.\ 
2004, \apjs, 154, 54 
\bibitem[Lutz et al.(2004)]{2004A&A...418..465L} Lutz, D., et~al.\ 2004, A\&A, 418, 465
\bibitem[Madau et al.(1996)]{1996MNRAS.283.1388M} Madau, P., et~al.\ 1996, \mnras, 283, 1388 
\bibitem[Magorrian et al.(1998)]{1998AJ....115.2285M} Magorrian, J., et 
al.\ 1998, \aj, 115, 2285 
\bibitem[Mart{\'{\i}}nez-Sansigre et al.(2007)]{2007MNRAS.379L...6M} 
Mart{\'{\i}}nez-Sansigre, A., et al.\ 2007, \mnras, 379, L6 
\bibitem[Men{\'e}ndez-Delmestre et al.(2007)]{2007ApJ...655L..65M} 
Men{\'e}ndez-Delmestre, K., et al.\ 2007, \apjl, 655, L65 
\bibitem[Mushotzky et al.(1993)]{1993ARA&A..31..717M} Mushotzky, R.~F., 
Done, C., \& Pounds, K.~A.\ 1993, \araa, 31, 717 
\bibitem[Oliver et al.(2000)]{2000MNRAS.316..749O} Oliver, S., et al.\ 
2000, \mnras, 316, 749
\bibitem[Page et al.(2004)]{2004ApJ...611L..85P} Page, M.~J., et al.\ 2004, \apjl, 611, L85 
\bibitem[P{\'e}rez-Gonz{\'a}lez et al.(2005)]{2005ApJ...630...82P} 
P{\'e}rez-Gonz{\'a}lez, P.~G., et al.\ 2005, \apj, 630, 82 
\bibitem[Polletta et al.(2006)]{2006ApJ...642..673P} Polletta, M.~d.~C., et 
al.\ 2006, \apj, 642, 673 
\bibitem[Polletta et al.(2007)]{2007ApJ...663...81P} Polletta, M., et al.\ 
2007, \apj, 663, 81 
\bibitem[Pope et al.(2008)]{2007arXiv0711.1553P} Pope, A., et al.\ 2008, 
ArXiv e-prints, 711, arXiv:0711.1553
%\bibitem[Simpson et al.(2006)]{2006MNRAS.372..741S} Simpson, C., et al.\ 
%2006, \mnras, 372, 741 
\bibitem[Smail et al.(2002)]{2002MNRAS.331..495S} Smail, I., et al.\ 2002, \mnras, 331, 495 
\bibitem[Stevens et al.(2004)]{2004ApJ...604L..17S} Stevens, J.~A., et al.\ 2004, \apjl, 604,
  L17 
\bibitem[Stevens et al.(2005)]{2005MNRAS.360..610S} Stevens, J.~A., et al.\ 2005, \mnras, 360, 610 
\bibitem[Swinbank et al.(2006)]{2006MNRAS.371..465S} Swinbank, A.~M., et al.\ 2006, \mnras, 371, 465 
\bibitem[Tremaine et al.(2002)]{2002ApJ...574..740T} Tremaine, S., et al.\ 
2002, \apj, 574, 740 
\bibitem[Valiante et al.(2007)]{2007ApJ...660.1060V} Valiante, E., et al.\ 
2007, \apj, 660, 1060 
\bibitem[Vignati et al.(1999)]{1999A&A...349L..57V} Vignati, P., et al.\ 
1999, A\&A, 349, L57 
\end{thebibliography}
\end{document}